\documentstyle[11pt]{article} 
\def\bas\def\baselinestretch{1.2} 
\begin{document} 
\begin{flushright}  
%{BONN-TH-29-P}\\   
\end{flushright}  
\vskip 2 cm 
\begin{center} 
{\Large {\bf Quadratic Divergences in  
Kaluza-Klein Theories}} 
\\[0pt] 
\bigskip {\large 
{\bf Dumitru M. Ghilencea\footnote{
{{ {\ {\ {\ E-mail: Dumitru@th.physik.uni-bonn.de}}}}}} } 
and {\bf Hans-Peter Nilles\footnote{
{{ {\ {\ {\ E-mail: Nilles@th.physik.uni-bonn.de}}}}}}
\bigskip }}\\[0pt] 
\vspace{0.23cm} 
{\it Physikalisches Institut der Universitat Bonn,} \\
{\it Nussallee 12, 53115 Bonn, Germany.}\\
\bigskip 
\vspace{3.4cm} Abstract
\end{center} 
{\small 
We investigate the so-called ``Kaluza-Klein regularisation''
procedure in supersymmetric extensions of the standard model with
additional compact dimensions and  Scherk-Schwarz mechanism 
for supersymmetry breaking. This procedure uses a specific 
mathematical manipulation to obtain  a finite result for the scalar
potential. By performing the full calculation, we show that the
finiteness of this result is not only a consequence of the underlying 
supersymmetry, but also the result of an implicit fine-tuning 
of the coefficients of the terms 
that control the ultraviolet behaviour. 
The finiteness of the Higgs mass at one-loop level seems therefore 
to be an artefact of the regularisation scheme, and quadratic
divergences are expected to reappear in higher orders of perturbation 
theory.
}
\newpage
\section{Introduction}

Perturbative calculations in quantum field theories are usually plagued
by the appearance of divergences. The program of regularisation and
renormalisation has shown, however, that  these
divergences do not lead to inconsistencies (in the absence of quantum
anomalies), but that they can be absorbed in a redefinition of the
physical input parameters. In the worst case (as for quartic and
quadratic divergences) one has to deal with a very strong dependence of
the parameters on the ultraviolet cut-off procedure. The knowledge of
physics in the far ultraviolet region has to be under control to have a 
meaningful 
understanding of the parameters of the theory even at 
very low energies. In
situations where various different mass scales appear one then
has to deal with the so-called hierarchy problems: why is one
scale small compared to the other? In the standard model of
particle physics this concerns the (mass)$^2$ of 
the Higgs boson (quadratically
divergent) and the vacuum energy (quartically divergent). The
cosmological constant and the Higgs mass are therefore naturally expected to be
of the order of the ultraviolet cutoff, e.g. the Planck scale. Of
course, it is not inconsistent to make an ad hoc choice of 
small values for these
quantities, but this would require an extreme fine-tuning of the input
parameters that has to be repeated order by order in perturbation
theory. Thus the presence of quartic and quadratic divergences appears 
highly undesirable from a theoretical point of view.

The above theoretical arguments were the main motivation to propose the
supersymmetric extension of the $SU(3)\times SU(2)\times U(1)$ 
standard model of
strong and electroweak interactions. Supersymmetric field theories are
free of quartic and quadratic divergences (provided some additional
constraints are fulfilled in the presence of abelian $U(1)$ gauge groups), 
as a result of a cancellation
between bosonic and fermionic degrees of freedom. 
In the calculation of
the vacuum energy e.g. we find that
the coefficient of the quartically divergent term is
proportional to the number of bosons minus 
the number of fermions while the coefficient of the
quadratically 
divergent piece is given by the supertrace of (mass)$^2$, STrM$^2$. 
Both of these vanish in the supersymmetric
limit. In fact, supersymmetry does even more than this 
due to the appearance of
so-called non-renormalisation theorems: corrections to the formerly
divergent quantities vanish identically in perturbation theory.  
In Nature however, supersymmetry is 
broken. Still, there exists the notion of softly broken supersymmetry (e.g.
spontaneous breakdown) where quadratic divergences for the Higgs mass are 
absent  and the highest degree of divergence (for field dependent
quantities) is logarithmic.
Quadratic divergences are cutoff due to the presence of new
physical states:
supersymmetric partners of the standard model particles. 
$M_{\rm susy}$, the supersymmetry breakdown scale (representative
for the mass of the supersymmetric partners) 
provides us with a
physical cutoff for the potentially quadratically divergent Higgs
(mass)$^2$, leading to a finite correction of order $M^2_{\rm susy}$. 
The quadratic divergence is recovered in the limit  
$M_{\rm susy}\rightarrow \infty$. 
Thus if supersymmetry is supposed to solve the
hierarchy problem, we expect supersymmetric partners in the 
TeV region (and not in the Planck mass region) to
provide this physical cutoff. Softly broken 
supersymmetry with $M_{\rm susy}$ of  the order
of the weak scale does therefore provide a solution to 
(at least the technical aspect of) the hierarchy
problem of the Higgs mass. In the present paper we adopt the definition
of soft supersymmetry breaking terms as defined in ref. \cite{palumbo}.
It will be these specific soft  terms we will have in mind
whenever  we will refer to soft breaking terms.
With these terms one can show that the
corresponding theory does not suffer from quadratic divergences
for scalar masses. This is true to all orders in perturbation
theory and is not just a reflection of  
the tuning of a one loop result.

So far our discussion only considered the presence 
of a {\it finite} number of fields. The situation
becomes much more complicated in the framework of string theories and/or
models of Kaluza Klein compactification where (from the D=4 dimensional 
field theory point of
view) we are dealing with infinitely many fields, and as a consequence
with infinitely many new ways to cutoff integrals and  fine-tune
parameters. In a given theory it is therefore important to control
the ultraviolet sensitivity of the parameters.
We have to understand whether a given value of
a physical parameter is the result of the specific choice of 
the cut-off or has a more general validity.

Some classes of string theories are believed to be finite in perturbation
theory. In the known cases, however, supersymmetry again plays a crucial
role in the discussion. A physical cutoff would therefore be
either the string tension (i.e. higher excitations of the string) or the
scale $M_{\rm susy}$ (i.e. supersymmetric partners).
In the present paper we shall not address the case of 
string theories, but the
simpler situation of compactified higher dimensional field theories
known under the name of Kaluza-Klein theories. Naively, such theories
show an ultraviolet behaviour worse than the corresponding 
four-dimensional $(D=4)$  theories
and some amount of supersymmetry is needed again 
to arrive at ultraviolet insensitivity.

The investigation of the present paper was initiated by claims in the
literature \cite{antoniadis,delgado,hall}
of an amazing ultraviolet softness of such Kaluza-Klein
theories. The  starting point is represented by 
supersymmetric theories in $D>4$  dimensions
(e.g. N=1 in  D=5) with specific hard (not soft
in the sense discussed above) breaking terms. From our
previous discussion we would therefore expect the appearance of e.g.
quadratic divergences for scalar mass terms. According to the claim of
 \cite{antoniadis,delgado,hall} these theories give a finite 
result. This claim is based on
the inspection of a one-loop contribution to the vacuum energy.
This in fact means that this theory has a better ultraviolet
behaviour than the underlying unbroken supersymmetric theory.

We find this a rather miraculous result that deserves further
clarification. After all we know that quadratic divergences 
can be regularised to give a finite result at one loop level and
the finite result will rely heavily on the details of the regularisation
procedure. 
The observed miraculous behaviour could thus be the consequence of a
special choice of the cut-off procedure. 
The calculational basis of the claims is a particular mathematical 
procedure in evaluating  the vacuum energy of the 
D=5 dimensional Kaluza-Klein
theory with hard supersymmetry breaking terms. The expression
of the vacuum energy is an infinite sum of 
infinite integrals  that are highly divergent.
The aforementioned procedure exchanges the sum and the integral to 
lead to a finite result (this regularisation procedure is
called ``KK-regularisation''). So far there has been no 
calculation that goes beyond the use of this procedure and  that would
clarify the detailed physical assumptions
employed in such a regularisation mechanism. It was 
argued \cite{antoniadis,delgado,hall}
that it must be the underlying (broken)
supersymmetry that leads to this result. Part of the
cancellation is in fact due to Supersymmetry. But
since the outcome of this calculation is a better ultraviolet behaviour 
than that of unbroken
supersymmetric theory, a more detailed calculation 
seems to be required.

In the present work we  perform such a calculation and we are able to 
clarify some of the points raised:
\begin{itemize}

\item The finite result requires a cancellation in bosonic 
(and fermionic) sectors
respectively (apart from other cancellations due to supersymmetry).

\item The regularisation procedure assumes 
a cancellation between physics above the momentum cutoff ($\Lambda$)
and  physics below this cutoff. 

\item The ultraviolet 
finiteness of vacuum energy  at one loop level 
appears as a result of a specific  fine-tuning of
parameters in the ``KK-regularisation''.

\end{itemize}

As a framework for our calculation and to motivate 
these  statements  we consider 
one particular example \cite{barbieri} of the models using
``KK regularisation''  and analyse this regularisation scheme which 
involves the summing up of the effects of an infinite tower of Kaluza 
Klein states associated with the extra (compact) dimension.
The model \cite{barbieri} we consider is representative for the class
of models considering ``KK regularisation'' 
\cite{antoniadis,delgado,hall,barbieri,quiros} 
and we only use it for illustrative purposes. 
To begin with we will only sum up the effects on the scalar potential of 
a finite number of  Kaluza Klein states
and then consider the limit when this number is infinite. This
approach has the advantage of showing rather explicitly  the origin
of the cancellation of the ultraviolet  divergences of the 
scalar potential. We argue that
there exists  such a  cancellation which takes place
separately for the bosons and for the  fermions, therefore is not 
related to the presence of supersymmetry (whether broken or not).
Rather we show that this is due to a subtle cancellation
between the quadratic terms of a finite number of  states in the Kaluza
Klein tower against the logarithmic contribution 
of such states of mass {\it larger} than the momentum cut-off of the 
one-loop integrals responsible for the radiative corrections to the scalar
potential. 
In a sense this can also be viewed as  a fine-tuning of the 
 coefficients
of the  ${\Lambda}$ dependent terms 
controlling the ultraviolet behaviour, 
such that the sum of these terms is zero when the number of Kaluza Klein 
states increases to infinity.
As usual (broken) supersymmetry does play  a role in the
cancellation of quartic and (even part of) the quadratic divergences, 
as a consequence of an equal number of bosonic and fermionic 
degrees of freedom.
But  as we just mentioned, the finite character of the scalar potential
(and soft masses) is traced back to individual cancellations
for bosons and for  fermions respectively due to
Kaluza Klein states of mass larger than $\Lambda$, the momentum
cut-off. Thus the  mechanism is not related to the 
way supersymmetry is broken. The finite character of the potential is
the result of a mathematical manipulation and is not necessarily 
supported  by an underlying physical mechanism or 
a symmetry principle.
 
In the next section we address the corrections to the scalar potential
of the model we investigate,
and refer the reader to \cite{barbieri} for a more detailed description.
In Section 3 we discuss the ultraviolet behaviour of the 
scalar potential. Our conclusions are presented in the last 
section. 

\section{The scalar potential}
The model \cite{barbieri} 
which provides the framework of our investigation considers a one-loop 
expansion of the  scalar potential,  whose radiative corrections
are accounted for by Kaluza Klein modes of the top quark,
associated with the extra spatial dimension considered. These states 
fall in N=2 multiplets, consequence of the enhanced supersymmetry in 
the 5D ``bulk'' (broken on the 4D boundary to N=0 by an orbifold-like
$S_1/{(Z_2\times Z_2')}$ compactification). 
The scalar  potential of \cite{barbieri} for the Higgs field 
$\phi$  is the focus of our investigation, has a rather 
generic structure with similar expressions found in 
\cite{antoniadis,delgado,hall}, and  is used to illustrate 
our analysis.
It has the structure 
\begin{equation}\label{potential1}
V(\phi)=\frac{1}{2} Tr \sum_{k=-l}^{l}\int \frac{d^4 p}{(2\pi)^4}
\ln\frac{p^2+m_{B_k}^2(\phi)}{p^2+m_{F_k}^2(\phi)}
\end{equation}
with the Trace over the top hypermultiplet of fixed k level
contributing a factor of $(4 N_c)$ where $N_c$ is the number of
colours.

In a 4D effective field theory approach, the tower of
Kaluza-Klein states should be truncated to a mass scale which 
should be of the order of the momentum cut-off scale of the loop
integral. 
This means that, at least intuitively, there should be a
correlation between the number of states in the tower and the momentum
cut-off. To keep our analysis general we first consider the 
case of an arbitrarily fixed number ($l$) of Kaluza-Klein modes,
which will prove very useful for gaining an insight into the physical 
meaning and  consequences of the regularisation procedure of  
summing over an infinite tower of Kaluza Klein states, as
explored  in \cite{antoniadis,delgado,hall,barbieri}.

The field dependent boson and fermion masses in (\ref{potential1})
are given by  
\begin{equation}\label{fermion}
m_{F_k}(\phi)=\frac{2 k}{R}+
m_t(\phi)=\frac{2}{R}(k+\omega),\,\,\,\,\,\,
\omega=\frac{m_t(\phi)R}{2}
\end{equation}
and
\begin{equation}\label{boson}
m_{B_k}(\phi)=\frac{2 k+1}{R}+
m_t(\phi)=\frac{2}{R}(k+\omega'),\,\,\,\,\,
\omega'=\omega+\frac{1}{2}
\end{equation}
and where following \cite{barbieri}  we consider any integer 
values of $k$ between $-\infty<k<\infty$. The use of these mass
formulas is again done to illustrate our analysis; other
similar possibilities \cite{antoniadis,delgado} may be explored, 
following this approach. The mass splitting 
between fermions and bosons, following the susy breaking via a
Scherk-Schwarz mechanism,  is therefore encoded in the
difference $\omega'-\omega$ which for this model is positive. 
The sign of the scalar potential
and also of  its second derivative  (at $\phi=0$) will crucially depend
on the sign of this difference (for this see eq.(\ref{potential1})), 
with implications for electroweak
symmetry breaking mechanism induced by radiative corrections.
Thus, the existence of this  mechanism is not really  a prediction of 
this class of models, but
can rather be traced back to the type of Scherk-Schwarz boundary 
conditions one  {\it chooses} for 
the fermions and for the bosons respectively, 
leading in some cases  to the presence
\cite{barbieri} or the absence \cite{antoniadis} of the electroweak 
symmetry breaking\footnote{This also depends on which zero 
modes we identify with the Standard Model particles.}.

The zero mode fermion in (\ref{fermion}) identified with the 
top quark has a mass given by \cite{barbieri}
\begin{equation}\label{top}
m_t(\phi)=\frac{2}{\pi R}\arctan{\frac{\pi R \phi y_t}{2}}
\end{equation}
This expression will not  be  used in 
our analysis and we present it only for illustrative purposes.

From equations (\ref{potential1}), (\ref{fermion}) and  (\ref{boson})
we find 
\begin{equation}\label{potential2}
V(\phi)=\frac{\eta}{R^4}
\sum_{k=-l}^{l}\int_{0}^{\overline\Lambda} {d \rho}\, 2\, \rho^3 
\ln\frac{\rho^2+\pi^2( k+\omega')^2}
{\rho^2+\pi^2( k+\omega)^2},\,\,\,\,\,\,\,\,\,\,
{\overline\Lambda}=\frac{\pi R \Lambda}{2}
\end{equation}
where the constant in front of $V$ is given by
\begin{equation}
\eta=\frac{4 N_c}{2 \pi^6}
\end{equation}
with $\Lambda$ as the momentum cut-off of the loop-integral 
in (\ref{potential1}).

Having truncated the tower of Kaluza-Klein states to a finite number,
one can safely commute the sum with the integral in
eq.(\ref{potential2}) without affecting its ultraviolet behaviour.
We therefore choose to keep two different cut-off's, one for the 
Kaluza-Klein sum (represented by $l$, see below)
and one for the loop-integral ($\Lambda$ or equivalently,
${\overline\Lambda}$). Strictly speaking
the number of Kaluza-Klein states should however 
be restricted to those states
whose mass is below the momentum cut-off of the loop integral,
condition which would lead to 
\begin{equation}
m_{B_k, F_k}\approx \frac{2 k}{R} \leq \Lambda \,\,\,\,\,\,
\Rightarrow \,\,\,\,\,\, {\overline\Lambda}\approx \pi l
\end{equation}
where $l$ stands for  the Kaluza Klein state of the largest 
mass. Later on we will consider the limit $l\rightarrow \infty$
with  $\Lambda$ (${\overline\Lambda}$) fixed  
to analyse the physical implications and mathematical subtleties
of the ``Kaluza-Klein regularisation''.

As a result of summing up the effects of a {\it finite} number of
Kaluza-Klein states  under the momentum integral  
in eq.(\ref{potential1}), (\ref{potential2}), we obtain a result with three
contributions under the loop-integral.
Accordingly, the potential can  be written as a sum of three 
terms, with the $\phi$ dependence hidden in $\omega$ and $\omega'$. Thus
\begin{equation}
V(\phi)=V_0(\phi)+V_1(\phi)+V_2(\phi)
\end{equation}
where we have
\begin{equation}\label{pot0}
V_0(\phi)= \frac{\eta}{R^4}
\int_{0}^{\overline\Lambda} {d \rho}\, 2\, \rho^3 
\ln\frac{\cosh(2 \rho)-\cos(2\pi\omega')}
{\cosh(2 \rho)-\cos(2\pi\omega)}
\end{equation}
and
\begin{equation}\label{pot1}
V_1(\phi)=\frac{\eta}{R^4}
\int_{0}^{\overline\Lambda} {d \rho}\, 2\, \rho^3 
\ln\frac{\rho^2+\pi^2( l+\omega')^2}
{\rho^2+\pi^2( l+\omega)^2}+(l \rightarrow -l)
\end{equation}
and finally
\begin{equation}\label{pot2}
V_2(\phi)= \frac{\eta}{R^4}
\int_{0}^{\overline\Lambda} {d \rho}\, 2\, \rho^3 
\ln\frac{[\Gamma(l \pm \omega'\pm i\rho/\pi)]_*}
{[\Gamma(l \pm \omega\pm i\rho/\pi)]_*}
\end{equation}
where under the last integral the symbol $[\Gamma(l \pm \omega'\pm
i\rho/\pi)]_*$ stands for a product of $\Gamma$'s with all 
possible (four) combinations of plus and minus signs. It is useful for
later reference to mention that the numerators ($\omega'$ dependent)
of the integrands of $V_0$, $V_1$, $V_2$ 
correspond to bosonic degrees of freedom, 
while the denominators  correspond to the contribution ($\omega$
dependent) of the fermions. Details of the exact calculation
leading to this result will be given elsewhere \cite{toappear}.

In the following we analyse each of the three contributions to
the scalar potential $V$ and investigate their physical implications. 
$V_0$ is part of the scalar potential corresponding to the result of 
reference \cite{barbieri} which is finite and ultraviolet insensitive,
as we discuss below. The expression of $V_0$ 
 simply corresponds to summing  over an infinite number of Kaluza
Klein states in eq.(\ref{potential1})  
and is therefore Kaluza-Klein-level independent. It was assumed
\cite{antoniadis,delgado,hall,barbieri} that this is the 
Kaluza-Klein regularised part of the scalar potential. 

Contributions $V_1$ and $V_2$ to the scalar potential 
are each   vanishing in the limit of an
infinite number of Kaluza-Klein states, $V_1(l\rightarrow \infty)=0$,
and $V_2(l\rightarrow \infty)=0$
while keeping $\overline\Lambda$ fixed, consequence of the vanishing
of their integrands respectively \cite{toappear}.  
It is however  instructive for our purposes to explicitly compute
$V_1$ and $V_2$  in the general case (of finite $l$)  
to understand the physical implications of taking
the limit $l\rightarrow \infty$ ($\overline\Lambda$ fixed).
This procedure will  also clarify whether 
the  breaking of supersymmetry by a Scherk-Schwarz-like
boundary condition  quoted  as  the origin of obtaining an  
ultraviolet finite result 
for $V$  in the limit $l\rightarrow \infty$,
plays actually any role in the cancellation of ultraviolet (quadratic and
logarithmic) terms of the scalar potential to give a finite 
result.

\section{Ultraviolet dependence of the scalar potentials}

To make our discussion more quantitative, we start with the analysis
of $V_0$. After performing the integral  \cite{barbieri} for $V_0$, 
one finds 
\begin{equation}\label{el}
V_0(\phi)= \frac{3\, \eta}{R^4}\sum_{k=0}^{\infty}
\frac{\cos(2 \pi\omega(2k+1))}
{(2k+1)^5}\left\{1-e^{-2{\overline\Lambda}(2k+1)}
\sum_{m=0}^{3}\frac{(2 {\overline\Lambda}(2k+1))^m}{m!}\right\}
\end{equation}
where the factor within curly braces is equal to unity in the large 
${\overline\Lambda}$ limit, to leave only the potential of 
\cite{barbieri}. Of the latter, only the first (small $k$) terms 
significantly contribute to  $V_0$.
In (\ref{el}) we used  $\omega'-\omega=1/2$, which
has implications for the sign of the potential (eq.(\ref{potential1})) 
and electroweak symmetry breaking as a result of the relation 
$\omega'>\omega$. Further analysis of this potential and its 
phenomenological implications have been discussed in 
\cite{barbieri}.

The result of integrating $V_1(\phi)$ of eq.(\ref{pot1}) 
is given, irrespective of the 
dependence $m_t(\phi)$, by the following expression
\begin{equation}\label{sum}
V_1=\underbrace{({\mathcal V}_1(\omega')+\Delta)}_{bosonic}
-\underbrace{({\mathcal V}_1(\omega)+\Delta)}_{fermionic}
={\mathcal V}_1(\omega')-{\mathcal V}_1(\omega)
\end{equation}
where we used the following notation 
\begin{equation}\label{bosonic}
{\mathcal V}_1(\omega')=\frac{\eta}{2 R^4}
\left\{\pi^2(l+\omega')^2{\overline\Lambda}^2
-\pi^4(l+\omega')^4\ln\left[1+\frac{{\overline\Lambda}^2}
{\pi^2(l+\omega')^2}\right]
+{\overline\Lambda}^4\ln\left(\pi^2(l+\omega')^2
+{\overline\Lambda}^2\right)
+(l\rightarrow -l)\right\}
\end{equation}
with terms obtained by the substitution $l\rightarrow -l$ to account
for modes moving in the opposite direction in the compact 
dimension.

Eq.(\ref{sum}) shows explicitly how some ultraviolet divergent terms
accounted for by $\Delta$ cancel
between the bosonic and fermionic contributions, to give $V_1$ as the
difference between a  (remaining) bosonic part which we denote by
${\mathcal V}_1(\omega')$
and  a fermionic one, ${\mathcal V}_1(\omega)$.
One sees for example  in the expression of $V_1$ 
how $\Lambda^4$ terms (included in $\Delta$)
 are cancelled due to the equal number of 
bosonic and fermionic degrees of freedom, consequence of the
initial presence of (now broken) supersymmetry.
Additional terms
${\overline\Lambda}^2$ and ${\overline\Lambda}^4
\ln(1+{\overline\Lambda}^2/\pi^2 (l\pm\omega')^2)$
are however present in $(\ref{bosonic})$ for the bosons 
with similar counterparts for the fermions\footnote{obtained by the
replacement $\omega'\rightarrow \omega$},
and their overall coefficient (at fixed $l$) is non-zero.
For example  ${\overline\Lambda}^2$ term has an
overall\footnote{including bosonic and fermionic contributions.} 
coefficient proportional to $(\omega'^2-\omega^2)$.
The absence of such terms in the final 
result for $V$ in the limit $l\rightarrow \infty$ has been 
assumed \cite{antoniadis,delgado,hall,barbieri}
to be due to the so-called ``soft nature'' of  breaking supersymmetry
by a Scherk-Schwarz mechanism which would preserve a cancellation between 
bosonic and fermionic contributions.
We argue that this seems not be the case, as terms  like  
${\overline\Lambda}^2$ and ${\overline\Lambda}^4
\ln(1+{\overline\Lambda}^2/\pi^2 (l\pm\omega')^2)$
do not cancel  exactly {\it between} the bosons and the fermions, for any 
value, whether finite or {\it infinite} of the number $l$
 of Kaluza Klein states that we summed over. 

To understand what really happens we consider  
the limit of  large $l$ and  fixed momentum cut-off 
$\Lambda$,  $\pi l\gg\overline\Lambda\equiv\pi\Lambda R/2$. In this limit
the logarithm in the  second term of the bosonic term 
${\mathcal V}_1(\omega')$ 
can be expanded in a (rapidly convergent) 
power series, with the first term in the expansion
to cancel the quadratic divergence ${\overline\Lambda}^2$ 
of the first term in ${\mathcal V}_1$; more explicitly, the second term 
in (\ref{bosonic}) which we denote as $A(l,\omega')$ has the form
\begin{equation}\label{expansion}
A(l,\omega')\approx 
\frac{1}{2}\frac{\eta}{R^4}\left\{
-\pi^2(l+\omega')^2 {\overline\Lambda}^2
+\frac{1}{2}{\overline\Lambda}^4-\frac{1}{3}
\frac{{\overline\Lambda}^6}{\pi^2(l+\omega')^2}+\cdots\right\}
\end{equation}
and its quadratic term\footnote{The  $\Lambda^4$ term in the expansion 
(\ref{expansion}) 
is cancelled in $V_1$ after including both bosonic and fermionic
contributions.} cancels the first term in ${\mathcal V}_1(\omega')$.
This  cancellation
is independent of the presence of (exact or softly broken)
supersymmetry, as  it takes place {\it separately} 
for the bosons and for
the fermions. The physical interpretation 
of this mathematical observation is that states 
of mass {\it larger} (!) than the momentum cut-off scale $\Lambda$ of the loop 
integral ($2 l/R \gg \Lambda$)
cancel the contributions to the quadratic divergence
of Kaluza-Klein states of level less or equal to $l$ and this
takes place separately for bosons and fermions, i.e. independent
of the presence/absence of supersymmetry or of the way it is broken.

Summing over an infinite tower of Kaluza Klein states
corresponds to the limit $l\rightarrow \infty$ 
and in this case one  finds that 
the quadratic dependence (${\overline\Lambda}^2$)
in the bosonic part
${\mathcal V}_1(\omega')$ of (\ref{bosonic})
is lost, and a similar mechanism separately applies for the fermionic part,
due to the aforementioned reasons, only to leave a
${\overline\Lambda}^4$ dependence cancelled between the bosonic and
fermionic parts. Indeed
\begin{equation}\label{v1limit}
\lim_{l\rightarrow \infty}
\frac{1}{2}\frac{\eta}{R^4}
\left\{\pi^2(l+\omega')^2{\overline\Lambda}^2
-\pi^4(l+\omega')^4\ln\left[1+\frac{{\overline\Lambda}^2}
{\pi^2(l+\omega')^2}\right]
+(l\rightarrow -l)\right\}=\frac{1}{2}\frac{\eta}{R^4}\,
{\overline\Lambda}^4
\end{equation}
where the r.h.s. 
${\overline\Lambda}^4$ is cancelled between bosons and fermions
due to the initial presence of supersymmetry.
The vanishing of ${\overline\Lambda}^2$ dependence can also 
be viewed as a fine-tuning of the 
($l$ and $\omega'$ dependent)  coefficients
of the two terms (momentum cut-off dependent) in eq.(\ref{v1limit})  
such that  their  sum only contains ${\overline\Lambda}^4$
but no quadratic contributions.
In the same limit $l\rightarrow \infty$, the third term in 
(\ref{bosonic}) ${\overline\Lambda}^4\ln(\pi^2 (l\pm\omega')^2+
{\overline\Lambda}^2)$ is indeed 
cancelled between bosons and fermions. This last cancellation
is in a sense  a remnant of the initial presence of supersymmetry 
which ensured the matching of bosonic and fermionic degrees of freedom,
and again brings in a fine-tuning to zero of the (logarithmic) 
coefficient of the 
ultraviolet term ${\overline\Lambda}^4 [\ln(\pi^2 (l\pm\omega')^2+
{\overline\Lambda}^2)-(\omega'\rightarrow \omega)]$ 
when $l\rightarrow \infty$.

One can argue that it is possible for the presence of quadratic 
ultraviolet terms in $V_1$ to be cancelled by similar terms which
may be present in $V_2$, to give - for any finite summation, up to
level $l$ -  a finite, ultraviolet insensitive result. 
Here we show that this is not the case by using 
an (approximate\footnote{The expression of $V_2$ (thus of $V$)
cannot be integrated
analytically to obtain a transparent result for our purposes
and this justifies the use of our approximation. 
In a different approach,
one can  first compute the integral of $V$ and then sum
over a finite tower of states. The summation is equally difficult to 
perform,  with the exception of that  for ${\overline\Lambda}^2$ terms, 
which  shows again that ${\overline\Lambda}^2$ terms are 
present for any finite summation, for both bosons and fermions and do not 
cancel between these. })
expression for $V_2$ valid in the limit of large modulus
of $l+i{\overline\Lambda}$ which 
does not restrict the relative values of $l$, ${\overline\Lambda}$, 
therefore various limits for these quantities may still 
be taken on the final result to find the ultraviolet behaviour of $V_2$.
After some algebra the integral of $V_2$ can be written as
\begin{eqnarray}\label{v2}
V_2& \approx & \frac{\eta}{R^4}
\left\{   \pi^2 g(l,\omega')(l+\omega')
\left[ {\overline\Lambda}^2-\pi^2(l+\omega')^2
\ln\left(1+\frac{ {\overline\Lambda}^2 }{\pi^2(l+\omega')^2}\right)
\right]
+(\omega'\rightarrow -\omega')\right\}\nonumber\\
&+&
\frac{\eta}{R^4}\frac{ \omega'{\overline\Lambda}^4}{2}
\left\{\ln\frac{{\overline\Lambda}^2+\pi^2( l+\omega')^2}
{{\overline\Lambda}^2+\pi^2( l-\omega')^2}\right\}
-\frac{\eta}{R^4}
\frac{4{\overline\Lambda}^5}{5 \pi}
\left\{\arctan\frac{\overline\Lambda}{\pi(l-\omega')}
+\arctan\frac{\overline\Lambda}{\pi(l+\omega')}
\right\}-(\omega'\rightarrow \omega)\nonumber\\
&+&
\frac{\eta}{R^4}\frac{ {\overline\Lambda}^4}{2}
 \left(l-\frac{1}{2}\right)
\ln\left\{
\frac{[{\overline\Lambda}^2+\pi^2( l\pm\omega')^2]_*}
{[{\overline\Lambda}^2+\pi^2( l\pm\omega)^2]_*}\right\}
\end{eqnarray}
where
$g(l,\omega')=[10-15
l+6l^2+3(-5\omega'+4l\omega'+2\omega'^2)]/60$
and  where the substitution 
$(\omega'\rightarrow\omega)$ only applies to terms in front of it,
to give the (separate, $\omega$ dependent) 
fermionic contribution. The last term in (\ref{v2})
contains both the fermionic and the bosonic contributions.
Ultraviolet divergences of the type ${\overline\Lambda}^2$,
${\overline\Lambda}^4$, ${\overline\Lambda}^5$ do cancel between the
bosonic and fermionic contribution while computing $V_2$, 
but these cancellations are {\it not} 
exact, with remaining (uncancelled) contributions 
(e.g. ${\overline\Lambda}^2$)
vanishing separately for the bosons and fermions, respectively.

In the first curly braces of (\ref{v2}) quadratic terms 
${\overline\Lambda}^2$ due to states up to level $l$ 
are cancelled for very large $l$ by the 
first term in the (rapidly converging) 
power series of the logarithmic contribution.
For $l\rightarrow \infty$ the whole (bosonic)
term in the first $\{\}$ has no ${\overline\Lambda}^2$
dependence,  giving only  a ${\overline\Lambda}^4$
contribution cancelled by its fermionic counterpart
due to the initial presence of supersymmetry.
The cancellation of ${\overline\Lambda}^2$ term  
takes place if the second term 
in the argument of the logarithm is smaller than unity, which is
only possible for Kaluza Klein mass terms {\it larger} than the 
momentum cut-off of the loop-integral\footnote{
Note that the sum of quadratic 
terms in $V_1+V_2$ does not vanish for any finite $l$.}.
This is true indeed in the limit $l\rightarrow \infty$
considered by  ``Kaluza Klein regularisation''.
A similar mechanism also applies to the case of the
fermionic contribution.

Other terms in (\ref{v2}) with ${\overline\Lambda}^{4,5}$ also vanish 
separately for the bosons and fermions respectively, 
in the limit $l\rightarrow \infty$, with fixed ${\overline\Lambda}$.
The vanishing in this limit
of such ultraviolet terms containing powers of $\Lambda$
brings in an amount of fine-tuning (to zero) of their ($l$ dependent) 
coefficients. This explains the  absence of these terms in $V_2$ (and
thus in $V$)
when $l\rightarrow \infty$.
To conclude, just as in the case of $V_1$, (part of) the 
 ultraviolet behaviour (e.g. ${\overline\Lambda}^2$)
of the bosonic (fermionic) part
is  separately  cancelled by bosonic 
(fermionic) contributions respectively, due to Kaluza-Klein 
terms of mass larger than the momentum cut-off of the loop
integral.

\section{Conclusions and outlook}
We have shown that the procedure 
of summing up an infinite tower of Kaluza Klein states
contributing to the scalar potential of supersymmetric models, 
used in the literature as a regularisation method
seems to have a rather unphysical meaning. 
Although some ultraviolet terms
do cancel between fermionic and bosonic contributions due to the initial
presence of supersymmetry, the cancellation of the (remaining) quadratic
and logarithmic divergences in this procedure to lead to an
ultraviolet finite potential
takes place independently for bosons
and fermions and is not a feature of  softly broken 
supersymmetry, as initially thought. 
The cancellation is simply due to considering Kaluza Klein 
states of masses larger than  the momentum cut-off
of the one-loop integrals for the bosons and for the fermions respectively, 
leading  us to a conclusion which questions the  physical meaning
of this regularisation procedure.

``Kaluza-Klein regularisation''  employs  an ad hoc choice for the 
cut-off of the quadratic divergence
at one loop order,  via a subtle mathematical procedure of summing over
an infinite number of Kaluza Klein states before performing the 
one-loop momentum integral. Even so, in contrast to the
situation in softly broken Susy, we will still have a strong ultraviolet
sensitivity of the Higgs mass since the quadratic dependence has been
cancelled at one loop only and will potentially reappear in higher orders
in perturbation theory.

Other unanswered questions by this string-inspired 
``KK regularisation'' 
come from  string theory. Unless a full string model is built to
justify  the summation over an infinite tower of states, 
one would expect the presence of a (unique)  cut-off in the theory
below which all masses are situated, and this was not the case
in the models addressed. In addition, 
any regularisation of string origin
should also  include the effects of 
possible winding modes with respect
to the compact extra dimension, and it is not clear at all that 
these will render the scalar potential ultraviolet insensitive
in a model embedded in a  full string set-up.
To exemplify, one can think of the heterotic string
calculation of string thresholds to the gauge couplings.
Such string threshold corrections may be viewed \cite{ferrara} 
as the free energy of compactification, thus intuitively related 
to the scalar potential in our model. 
These corrections still exhibit  (for  two-torus compactification)
a quadratic dependence on  the compactification scale 
 as well as on the string  scale \cite{dixon,ferrara}.  
For  an effective field theory model the  string scale 
is assimilated to the 
ultraviolet cut-off, thus an ultraviolet behaviour  of the potential 
stronger than the (ultraviolet insensitive)  finite 
one of the  ``KK regularisation''  should then be  expected.

\vspace{0.5cm}
\noindent {\bf Acknowledgement} 
We  would like to thank Graham G. Ross and  
Stefan Groot Nibbelink for very useful discussions on this work.
This work  was supported by the
University of Bonn under the European Commission RTN programmes 
HPRN-CT-2000-00131 and 00148.

\end{document}